%% file: authordraft_IEEE.tex
\documentclass[conference]{IEEEtran}
\IEEEoverridecommandlockouts

\usepackage{cite}
\usepackage{amsmath,amssymb,amsfonts}
\usepackage{algorithmic}
\usepackage{graphicx}
\usepackage{textcomp}
\usepackage{xcolor}
\usepackage{caption}
\usepackage{subcaption}
\usepackage{hyperref}

\usepackage{flushend}
\usepackage{fancyhdr}
\def\BibTeX{{\rm B\kern-.05em{\sc i\kern-.025em b}\kern-.08em
    T\kern-.1667em\lower.7ex\hbox{E}\kern-.125emX}}

\begin{document}

\title{Learning Intermediate Representations using Graph Neural Networks for NUMA and Prefetchers Optimization}

\author{\IEEEauthorblockN{Ali TehraniJamsaz\textsuperscript{1}, Mihail Popov\textsuperscript{2}, Akash Dutta\textsuperscript{1}, Emmanuelle Saillard\textsuperscript{2}, Ali Jannesari\textsuperscript{1}}
\IEEEauthorblockA{\textsuperscript{1}\textit{Iowa State University, Ames, Iowa, USA}}
\IEEEauthorblockA{\textsuperscript{2}\textit{Inria, Bordeaux, France}}
\IEEEauthorblockA{\{tehrani, adutta, jannesar\}@iastate.edu}
\IEEEauthorblockA{\{mihail.popov, emmanuelle.saillard\}@inria.fr}
\\[-3.0ex]
}


\maketitle

\input{0_abstract}


\input{1_intro}

\input{2_background}

\input{3_approach}

\input{4_result}

\input{5_related_works}

\input{6_conclusion}

\bibliographystyle{IEEEtran}
\bibliography{sample-base}

\end{document}

%% file: 0_abstract.tex
\begin{abstract}
There is a large space of NUMA and hardware prefetcher configurations that can significantly impact the performance of an application. Previous studies have demonstrated how a model can automatically select configurations based on the dynamic properties of the code to achieve speedups.
This paper demonstrates how the static Intermediate Representation (IR) of the code can guide NUMA/prefetcher optimizations without the prohibitive cost of performance profiling. 
We propose a method to create a comprehensive dataset that includes a diverse set of intermediate representations along with optimum configurations. We then apply a graph neural network model in order to validate this dataset. 
We show that our static intermediate representation based model achieves 80\% of the performance gains provided by expensive dynamic performance profiling based strategies.
We further develop a hybrid model that uses both static and dynamic information. Our hybrid model achieves the same gains as the dynamic models but at a reduced cost by only profiling 30\% of the programs.
\end{abstract}

%% file: 1_intro.tex
\section{Introduction}
Thread-level parallelism increases data demand to the main centralized memory. Complex memory hierarchies, caches, and prefetching methods that identify access patterns and fetch the data ahead of time try to solve this but have not scaled with the data demand, thus limiting the performance gains. To continuously sustain memory scaling, multi-core systems were forced to adopt the Non-Uniform Memory Access (NUMA) memory design. NUMA systems provide a separate memory for each processor, thereby increasing overall bandwidth and reducing congestion through the centralized memory. 

The resulting complex hardware exposes a large parameter space that must be explored for efficient application execution. 
Many tuning strategies and heuristics have been implemented to explore  NUMA threads~\cite{popov2017piecewise,durillo2018multi,wang2016predicting} and pages~\cite{diener2015locality,majo2012matching,dashti2013traffic}, as well as prefetching~\cite{wu2011characterization,khan2015arep,jimenez2012making}. A common method is to build models on performance counters~\cite{liao2009machine,Denoyelle:2019:DTP:3337821.3337893,10.1145/3392717.3392765}. While executing the programs, performance metrics, that describe the execution behavior, are collected to guide the choice of the optimization online~\cite{dashti2013traffic,broquedis2010forestgomp} or offline\cite{diener2015characterizing,trahay2018numamma}. For instance, if we observe many remote accesses from a single node, it can be efficient to scatter the pages to the rest of the system. 


Models either rely on expert knowledge or on automatic machine learning approaches \cite{rameshka2019rigel, milani2020autotuning}. Analytical models are understood by researchers but are difficult to adapt to the constant ongoing evolution of software and hardware. As a result more and more Machine Learning (ML) models are emerging\cite{Denoyelle:2019:DTP:3337821.3337893}.
There is a trade-off between the overhead and the complexity of the ML model and the output prediction accuracy. More advanced techniques such as deep learning can produce better predictions but require significantly more data. As a result, deep learning frameworks are not currently applied to guide NUMA/prefetching due to both the search size and the prohibitive cost of performance profiling \cite{10.1145/3392717.3392765}. 

Indeed, collecting performance counters
is very time consuming, preventing the collection of thousands of samples required for deep learning. Even worse, some hardware systems have custom performance counters. In other words, models targeting one system are not portable to another system as they have different sets of performance counters. 


This paper proposes a novel code characterization method that avoids the prohibitive cost of dynamically profiling applications while providing valuable information that can be used to guide the optimization of programs with deep learning strategies. Our method statically collects data and thus enables the usage of advanced deep learning strategies to deliver competitive results.
Moreover, our approach automatically identifies cases where only using static information is sufficient for optimization, and thus only collects dynamic information if needed.

We start by compiling applications with different flag sequences. Disparate compiler optimizations expose different properties of the code. For instance, the \textit{dead code elimination} compiler pass removes useless blocks in the application. If a code has large blocks of useless code, this compiler pass will have a significant impact. The key idea is to quantify how a compiler transformation affects a code and feed this information to a deep learning framework, i.e., we leverage the fact that compiler transformations impact codes based on their properties to guide the overall optimization process. 

We use existing Intermediate Representation (IR) embedding techniques\cite{cummins2020programl} to quantify the impact of a compiler pass\cite{lattner2004llvm}. The intermediate representation of a code can be represented by a graph which is used by graph neural networks (GNN) as input to produce a vector as output. We expect that similar codes will have similar vector representations for the GNN outputs.

Once the programs have been compiled with custom flag sequences into IR and subsequently graphs, we feed a graph-based deep learning framework with the programs' graphs along with a best configuration to use for the programs. We note that the best configuration is only defined by exploring the space of configuration once with the default compiler settings. The resulting deep learning model predicts the best configuration for a new unknown application with its default compiler settings. The model just needs to recompile the application with a custom flag sequence selected either by a model or by simply picking the most efficient one at training. 

We demonstrate the capabilities of our characterization technique by training a deep learning framework over NUMA and prefetch configurations considering two hardware micro-architectures: Intel Sandy Bridge and Intel Skylake. 

While operating only on static information, our model is able to achieve 80\% of gains provided by dynamic approaches. We further show how our model is able to predict configurations across different micro-architectures by training it for one micro-architecture and applying it to another (e.g., trained on Sandy Bridge and applied to Skylake).
We also illustrate how a hybrid approach consisting of both static and dynamic information can help to achieve the same performance gains as dynamic approaches while only profiling 30\% of the codes. 

Overall, the contributions of this paper are:

\begin{itemize}
    \item A new method of constructing a dataset by applying numerous compiler transformation flags;
    \item A novel static code characterization method that compiles the same code across different flag sequences to expose its properties;
    \item An application of the proposed method over a complex tuning parameter search space of NUMA and prefetch configurations;
    \item A novel GNN-based hybrid model that only requires dynamic information when static information is not sufficient;
    \item A quantification of the  performance portability losses when we train and apply models across different micro-architectures.
\end{itemize}

This paper is structured as follows: Section \ref{background} provides the background while Section \ref{methodology} illustrates our approach by explaining how we create the dataset and use it with graph neural network model to learn the characteristics of programs. 
Section \ref{resutls} presents the experiments. Section \ref{sec:disc} discusses limitations and future work. 
Section \ref{related} presents the related works in prefetching and NUMA optimizations, and Section~\ref{conclusion} concludes the paper. 

%% file: 2_background.tex
\begin{figure*}
\centering
\includegraphics[scale=0.58]{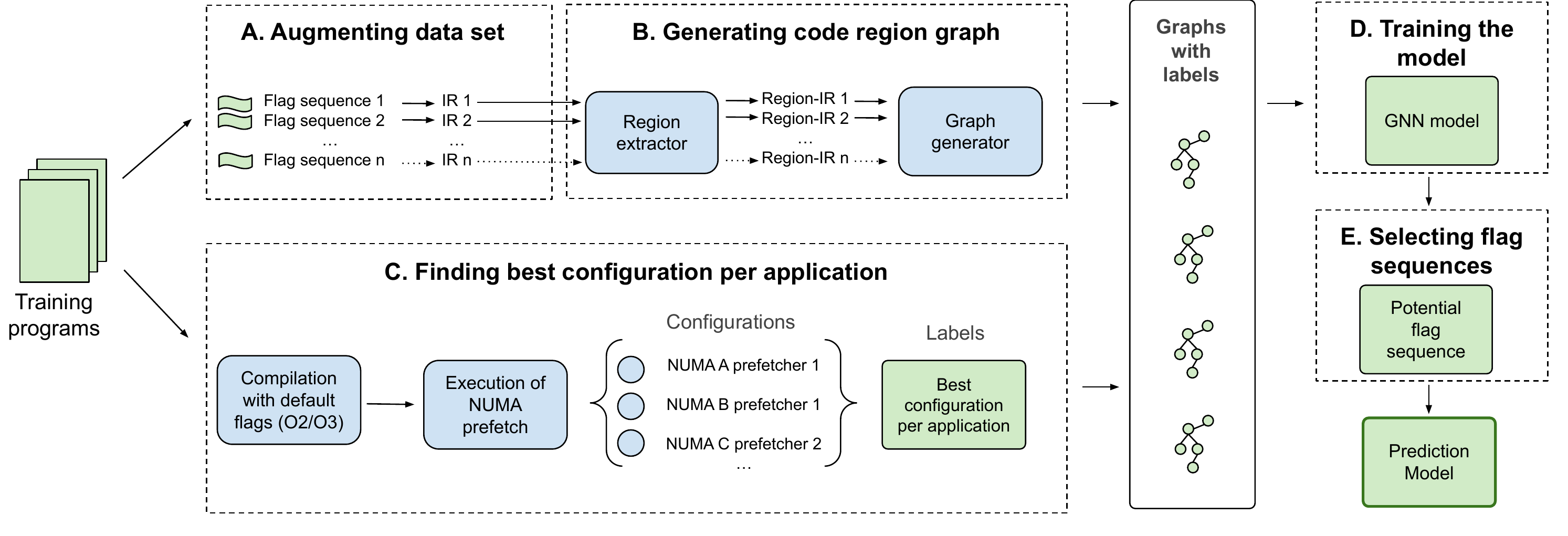}
\caption{Workflow of training a GNN model to predict NUMA/Prefetching optimizations based on static information.}
\label{fig:training}
\end{figure*}

\section{Background}
\label{background}

This section describes background knowledge for programs' representations, graph neural networks, and NUMA and prefetching optimizations.

\subsection{Graph representation of programs}
Performance counters are commonly used to represent applications \cite{10.1145/3392717.3392765,liao2009machine, alcaraz2021building}. One drawback of this representation is its cost. We need to profile the application by executing it. Moreover, this representation is hardware specific, thus limiting the cross-architecture characterization. 
An alternative to avoid these issues is to apply natural language techniques on the source code to represent the applications \cite{allamanis2016convolutional, raychev2014code, tehranijamsaz2021deeprace}. They mainly consider source code as a sequence of tokens. However, this representation ignores the rich structural information that can be essential for a model to characterize an application.

Recent studies \cite{ben2018neural, allamanis2017learning, raychev2016probabilistic, maddison2014structured} include this missing structural information from the source code, by representing the applications as graph structures.
Different types of graphs are used. For example, Nair et al.\cite{nair2020funcgnn} use the control flow graph of each program in order to learn if two programs are similar.
Allamanis et al.\cite{allamanis2017learning} augment the Abstract Syntax Tree (AST) of programs by introducing additional edges in the AST, and show that with augmented AST, a Graph Neural Network can predict variable names and variable misuses.

Recently, Cummins et al.\cite{cummins2021programl} proposed a comprehensive graph representation called ProGraML based on the Intermediate Representation (IR) of the programs. Unlike most approaches which rely on control flow or AST, they construct a graph with three different flows: control flow, data flow and call flow. Their proposed graph representation sets new state-of-the-art results on a number of downstream tasks like algorithm classification.
Given the comprehensiveness of their graph representation, we specifically reuse their approach in order to generate graphs for programs in our dataset.

\subsection{Graph Neural Networks}

Deep leaning models such as Convolutional Neural Networks and Recurrent Neural Networks have shown great success in areas of image processing\cite{sultana2018advancements} and natural language processing\cite{chen2016enhanced}.
Recent work\cite{allamanis2017learning} shows that neural networks can efficiently learn the structural and semantic information of source code when programs are represented as graphs.
Graphs carry two distinct types of information: nodes and edges. As a result, a new type of neural networks called Graph Neural Networks\cite{zhou2020graph} (GNN) capable of processing graph data structures has emerged.

Almost all GNNs are implemented using the Message Passing Neural Networks\cite{gilmer2017neural} (MPNN) framework.
 The goal of these networks is to learn the latent space representation of each node through its neighbouring nodes in the graph. To this end, there are three important functions that construct a MPNN:
 \begin{enumerate}
  \item \textit{Message:} constructs the message which exchanges along the edges between the target node and its neighbouring nodes.
  \item \textit{Aggregate:} aggregates all messages that are received from neighbouring nodes.
  \item \textit{Update:} Updates the target node embedding according to Message and Aggregate functions. 
\end{enumerate}

Relational Graph Convolution Network\cite{schlichtkrull2018modeling} (RGCN) are a type of graph neural network capable of updating the embedding state of each node considering the type of relations (edges) between the target node and neighbouring nodes.
Equation \ref{RGCN_update_formula} shows how a latent representation of a node is updated in a RGCN.

\begin{equation}
    h^{l+1}_i=\sigma{(W_0^lh_i^l + \sum_{r \in R}\sum_{j \in N_i^r}\frac{1}{c_{i,r}}W_r^lh_j^l)}
\label{RGCN_update_formula}
\end{equation}

Where $h_i^{l+1}$ is the latent representation of node $i$ in $(l+1)^{th}$ layer. $\sigma$ is the update function. The aggregation function is represented by the two summations. $W_r^l$ is the weights for relation type $r$ for layer $l$ and $C_i$ is a normalization constant.
We use a RGCN along the previously described graph representation as it can efficiently learn the different types of edges (flows) in the graphs.

\subsection{NUMA and Prefetching} 

Optimizing NUMA and prefetch requires to explore a large search space. This paper focuses on the space parameterized by Sánchez Barrera\cite{10.1145/3392717.3392765}. In particular, we focus on the hardware prefetcher configurations provided by Intel. Each core is equipped with four independent prefetchers which can be enabled or disabled by simply updating a Model-Specific Register (MSR) \textit{0x1A4}\cite{Hiebel:2019}. The prefetchers include a \textit{Data cache Unit (DCU) IP-correlated} that correlates prefetches with the Instruction Pointer (IP), a \textit{DCU prefetcher} that brings the next L1 cache line, an \textit{L2 adjacent cache line prefetcher}, and a \textit{L2 streamer prefetcher}. We explore the 16 possible different combinations of prefetching for each workload. 

We couple the 16 combinations of prefetchers together with the optimization space of NUMA originally described by Popov et al.\cite{popov2019efficient} which includes degree of parallelism, degree of NUMA node, thread mapping (i.e., round robin, contiguous), page mapping (i.e., First touch, locality, interleave, balance) resulting in 288 or 320 configurations to evaluate depending on the system (Skylake or Sandy Bridge). While large, this optimization space provides an arithmetic average speedup over $2\times$ against an already optimized default (uses all cores and NUMA nodes, data locality, all prefetchers on).

Finally, Sánchez Barrera et al.\cite{10.1145/3392717.3392765} reported that exploring 13 selected configurations instead of 288/320 provides 99\% of the performance gains achieved by fully exploring the search space. We use these 13 configurations as labels for our models. They significantly simplify the training while providing similar gains as the full space.

%% file: 3_approach.tex

\section{Approach}
\label{methodology}

S\'{a}nchez Barrera et al.~\cite{10.1145/3392717.3392765} show that machine learning models can automatically optimize NUMA and prefetching configurations based on the dynamic properties of the program. However, obtaining such dynamic properties for an application comes at the expensive cost of executing the application. In this section, we present an approach to automatically select a configuration by only using the static properties of the program. Then, we extend it with a classifier that determines whether or not the static information is enough to optimize the application.


Figure \ref{fig:training} presents the prediction workflow for NUMA/prefetch configuration from static information. 
Given a set of programs (referred as \emph{Training programs}), step A augments the training data by applying various flag sequences. Then, regions of interests are extracted and graph representations of those regions are created in step B.
In step C, the best configuration per application is identified.
Step D uses the graphs and the identified configurations to train a graph neural network-based model to predict the best configuration.
In step E, we select the flag sequence to use when characterizing new unseen applications. We propose two approaches based on the training programs. We either review the flag sequences to identify the sequence that leads to the best average predictions or we train a static model that predicts the most efficient flag sequence by characterizing the unseen program. 

Finally, we further enhance the prediction by incorporating another model which can identify the cases where static information is not sufficient. In that case, dynamic information (i.e., performance counters) will be collected.
Each step of the workflow are explained in details in the following subsections.

\subsection{Augmenting Dataset (step A)}
The number of samples in a dataset plays a vital role in training deep learning models to achieve high accuracy.
Unfortunately, HPC benchmark suites are generally not large enough to train such models.

Inspired by data augmentation techniques in image processing field~\cite{paulin2014transformation} (e.g., flipping) as well as from the compiler community when optimizing code size\cite{da2021anghabench}, we propose to transform the representation of the code.
To this end, we leverage various compiler optimization flags offered by the LLVM compiler\cite{lattner2004llvm} to produce different forms of the same program. 
This augmentation technique helps to alleviate the problem of insufficient training samples. We compile the source code files with specific compiler sequences 
to create numerous Intermediate Representation files.


To create middle-end flag sequences, we followed the methodology described in ~\cite{popov2017piecewise} where random compilation sequences are generated by down-sampling the \textit{-O3} sequence.  
Each pass is removed with a 0.8 probability and the process was repeated four times. Our goal is not to find the flag sequence that optimizes the code, but instead to explore diverse optimizations that are used to better understand the code.




\subsection{Generating Code Region Graph (step B)}

\textbf{Code Region Extraction:} 
Programs can be statically analyzed at different levels of granularity: compilation units, functions, loops, instructions. However analyzing unrelated instructions introduces noises, which may result in a lower prediction accuracy.
This work optimizes OpenMP regions. In LLVM, parallel regions are implemented as outlined functions in the IR\cite{popov2015pcere}.
Therefore, we extract these functions with \texttt{llvm-extract} tool into small stand alone IR files. 


\begin{figure}%
    \centering
    \subfloat[\centering Static model]{{\includegraphics[scale=0.6]{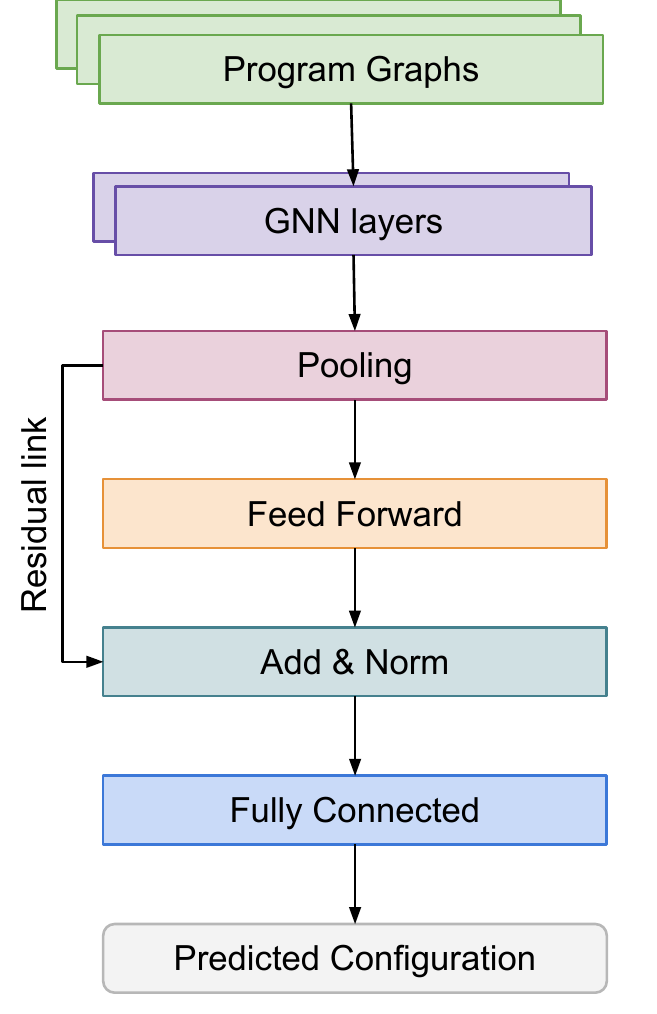} }}%
    \qquad
    \subfloat[\centering Hybrid model]{{\includegraphics[scale=0.6]{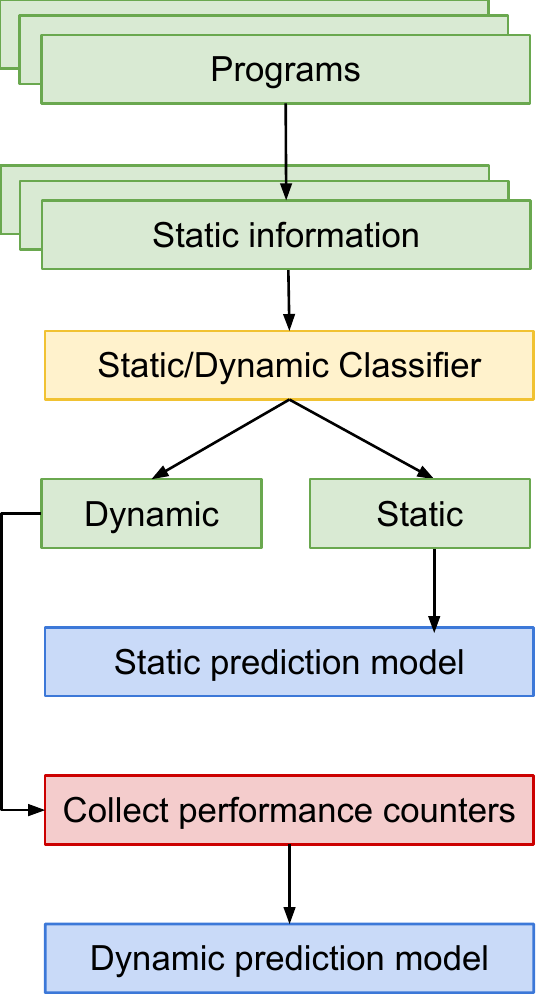} }}%
    \caption{Prediction model architecture}%
    \label{fig:training_model}%
\end{figure}

\textbf{Program representation:} 
LLVM middle-end passes generate Intermediate Representations that can be represented as graphs \cite{cummins2020programl}. We convert each file into a graph to expose it to the GNN. 




\begin{figure*}[h]
\centering
\includegraphics[width=\textwidth]{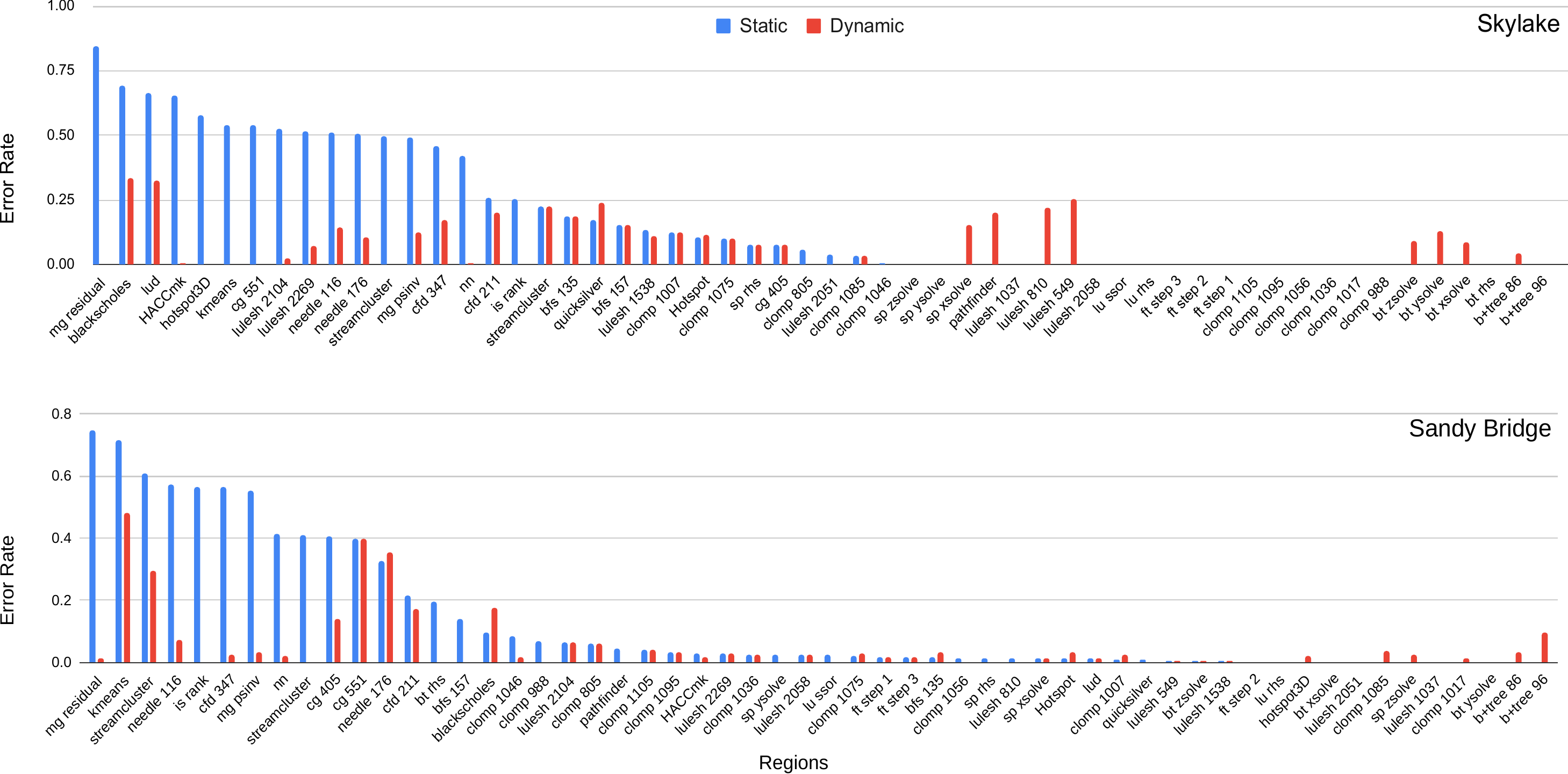}
        \caption{[lower is better] Breakdown of the prediction errors per region.
For each region, we compare the prediction results of the static model using the explored flag seq method against the dynamic performance counters based model.}
        \label{errorperregion}
\end{figure*}

\subsection{Finding Best Configuration per Application (step C)}
In parallel to steps A and B, step C assigns to each application the configuration label that provides the best performance for that application. To do so, we compile the applications with their default flags (e.g., \textit{O2/O3}) and then execute once on the whole NUMA and prefetch space to collect the timings. This is a necessary step and unavoidable cost to identify the labels that we pay only once; we can then apply the static model without any dynamic execution.


\subsection{Training the Prediction Models (step D)}
\label{modelhh}
Once the graphs with labels are prepared for training, a graph neural network model is deployed over the programs' graphs to learn their characteristics. 
As discussed in section \ref{background} we construct a graph neural network model based on the RGCN layers.
Figure \ref{fig:training_model} shows the architecture of the proposed prediction models on top of graph neural networks. In particular, we propose two models. One model predicts the configuration of a program statically, while the other one identifies cases where the static prediction model fails in its prediction. Thus the second model delegates the configuration prediction to a dynamic model if needed.
 
Each program graph representation is passed through different layers such as RGCN, Pooling, and Fully Connected Layers. We also utilize a residual connection\cite{he2016deep} followed by a normalization layer.
The output of the normalization layer is a vector for each program's graph.
 
 \subsubsection{Static model}
 The vectors retrieved after the normalization layer are used as inputs to a fully connected neural network (FCNN) (Figure \ref{fig:training_model} (a)). 
 This FCNN receives the vectors as input and tries to predict the correct configuration based on the input vector.
Training this model enables the prediction of best configurations based on static features only. Section \ref{sec:exp:stat} presents the results of the static model.
 
 \subsubsection{Hybrid model}

Once the static model is deployed, we use the error rate (difference between the full exploration performance gains and the predicted performance gains) in its predictions as labels to train another model (shown as Hybrid model in Figure \ref{fig:training_model} (b)). This hybrid model identifies whether the static model is able to predict the configuration of a new unseen application or if it is better to delegate the task of configuration prediction to a dynamic model which depends on performance counters. In such cases, the application is  further profiled to collect the relevant performance counters~\cite{10.1145/3392717.3392765}.

The hybrid model uses the statically generated vectors from the normalization layer as inputs. We fed these vectors along with the errors labels to a decision tree using the default Scikit-learn setup\cite{pedregosa2011scikit}. We additionally use Genetic Algorithms (GA) implemented with \textit{pyeasyga} as a feature reduction technique to subset the vectors. The original vector size is 256. We explore through GA different combinations of vectors subsets of 10 element. This improves the model accuracy by removing parts of the vector that contains noisy information. The GA uses a population size of 500, a crossover of 0.8, and a mutation rate of 0.1. Section \ref{sec:exp:hyb} presents the results of the hybrid model.

\subsection{Selecting Flag Sequences (step E)}
\label{sec:stepe}
We use many different flag sequences to augment the dataset and train our models. However, we do not want the end user to recompile their application multiple times to apply this method. Therefore, we need to select a single flag sequence to use at deployment.

We develop two methods to select this flag sequence at training. First, once the static optimization model has been completely trained over all the training programs, we re-evaluate all flag sequences and look at their average predicted performance across all training programs. We return the sequence that provides the best results. While simple and efficient, this approach provides sub-optimal performance gains for unseen programs that require different flag sequences to be characterized. Indeed, a static model always uses the same flag sequence for all unseen programs. 

On the opposite, the second method is a flag prediction model that selects a flag sequence based on the program properties. This enables to use different flag sequences, directly adjusting them for each program. While this approach is more efficient, it further complicates the optimization process. The flag prediction model uses exactly the same learning algorithm and features as the hybrid predictor. Static vectors are generated using a fixed flag sequence, then fed to a decision tree and explored by a GA to reason about how to sample the vectors.
The difference between the two models is the labeling. Instead of predicting the static errors, the flag prediction model predicts compiler sequences.

We call \textit{explored flag seq} and \textit{predicted flag seq} the sequences selected by exploring the sequences over the training programs and by using the flag model respectively. Section \ref{sec:exp:stat} quantifies the impact of selecting flag sequences while Section \ref{sec:res:modelflag} presents the gains of predicting flag sequences.

%% file: 4_result.tex
\section{Experimental Results}
\label{resutls}
In this section, first we evaluate our static prediction model.
We further investigate how to improve the model accuracy and illustrate how graph-based deep learning models operate across different contexts such as cross-architectures or cross-inputs.
Finally, we evaluate the performance of the hybrid approach and quantify the gains of using models to select flag sequences. 




\subsection{Experimental Setup}

To evaluate the prediction of our approach, we directly compare our results to a dynamically based solution~\cite{10.1145/3392717.3392765}. To the best of our knowledge, in~\cite{10.1145/3392717.3392765}  only their reaction based performance counters models have successfully optimized the coupled search space of NUMA and prefetchers. We refer to their most efficient model (i.e. classification tree using power package and L3 miss ratio) as \textit{dynamic} and use the same environment to compare it against our static solutions.

In particular, we used the same OpenMP parallel regions from Rodinia~\cite{RodiniaPaper} and the NAS C Parallel Benchmarks~\cite{NPBPaper,popov2015pcere} version~3.0, along with LULESH version~2~\cite{LULESH:spec} and CLOMP~\cite{bronevetsky2008clomp} executed with the same (except if stated otherwise) LLVM Clang version~3.8~\cite{lattner2004llvm} over a four-node Intel Sandy Bridge EP E5 4650 and a dual-node Intel Skylake Platinum 8168. We reuse the already profiled timings and performance counters for all the 57 regions except one. \textit{IS random generator} resulted in missing compilation data for our model so we removed it. 

To validate all our model types (i.e., static, hybrid, and flag prediction), we both decompose the programs into the same 10-folds for cross validation and apply the same 13 configurations, which preserve the performance gains of the whole space while drastically reducing its size (see Section \ref{background}), for labeling as Sánchez Barrera et al.~\cite{10.1145/3392717.3392765}. Because of cross-validation, for each type of model, we train 10 different instances and evaluate each one over a validation fold composed of approximately 5 new unseen programs using the remaining 9 folds for training. This ensures that our models can operate over unseen programs.

All the prediction numbers/characteristics presented in the experiments are generated by aggregating the programs from all the validation folds. We note that all speedups in the paper are calculated against the already optimized configuration; data: locality, threads: scatter, prefetchers: all enabled, where all cores and NUMA nodes are used.


\begin{figure}[h]
\centering
     \begin{subfigure}[Sandy error distribution]{0.5\textwidth}
         \centering
         \includegraphics[width=\textwidth]{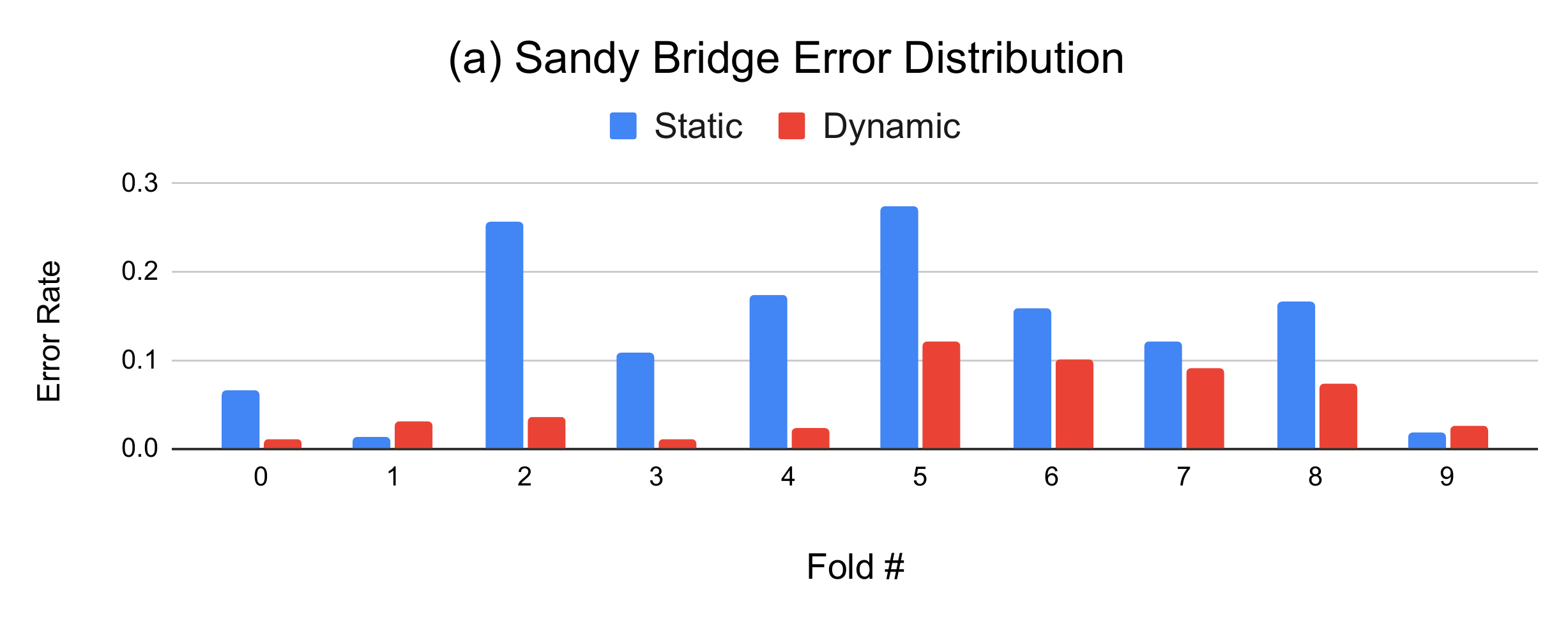}
     \end{subfigure}
     \hfill
     \begin{subfigure}[Skylake error distribution]{0.5\textwidth}
        \centering
         \includegraphics[width=\textwidth]{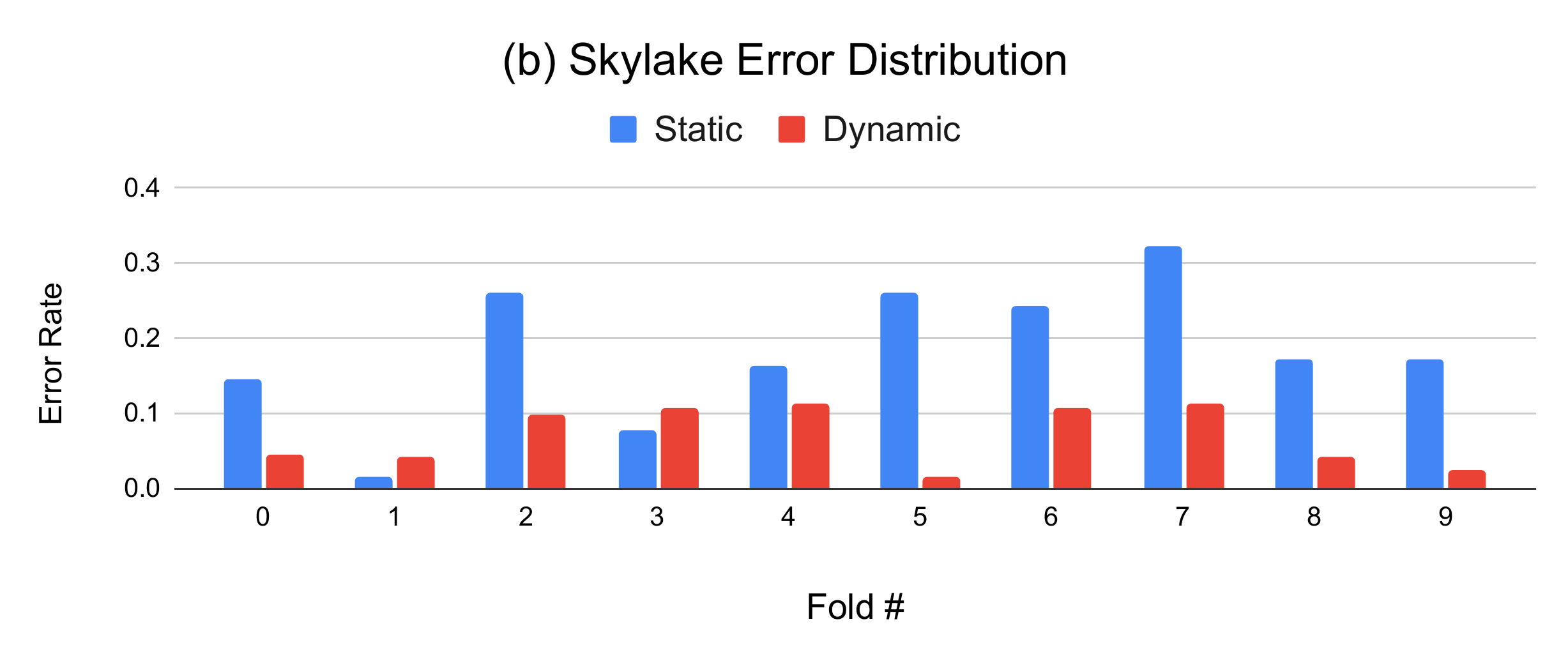}
     \end{subfigure}
        \caption{[lower is better] Breakdown of the average prediction errors (calculated with relative differences) per validation fold. We observe relatively consistent errors across the folds.}
        \label{errorfold}
\end{figure}

\begin{figure}[h]
\centering
     \begin{subfigure}[skylake and skylake prediction]{0.5\textwidth}
        \centering
          \includegraphics[width=\textwidth]{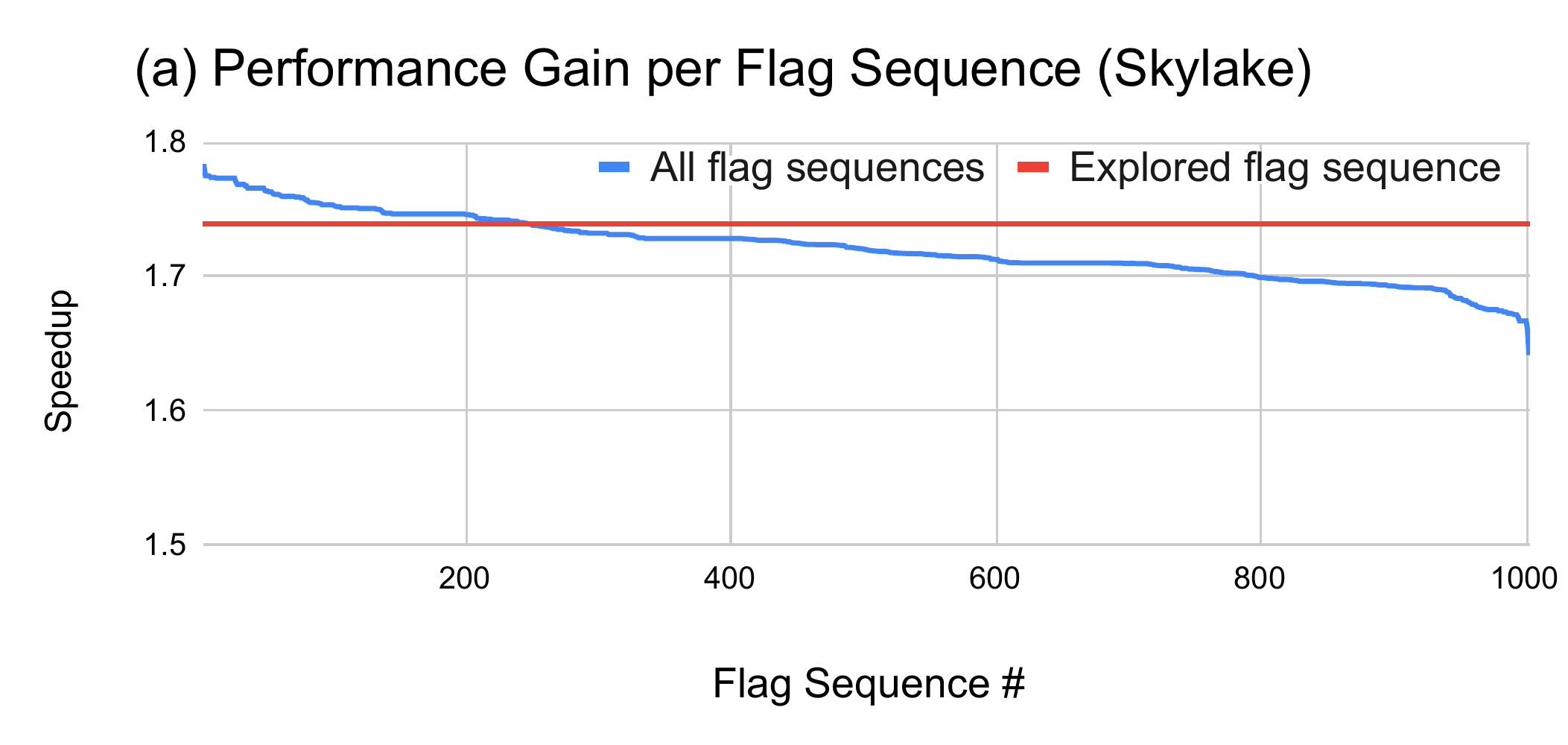}
     \end{subfigure}
     \hfill
     \begin{subfigure}[sandy and sandy prediction]{0.5\textwidth}
        \centering
         \includegraphics[width=\textwidth]{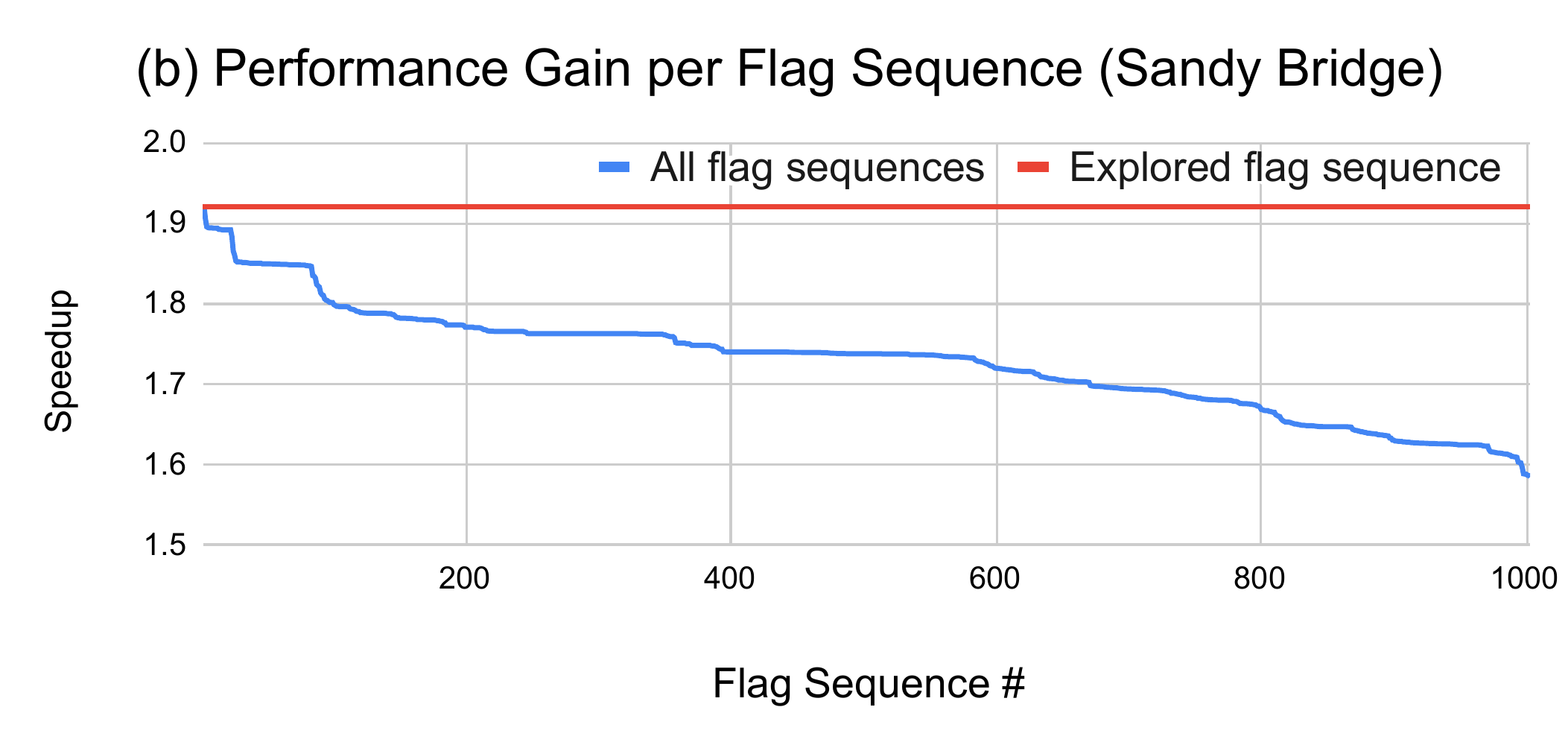}
     \end{subfigure}
        \caption{[higher is better] Arithmetic average speedup achieved per flag sequence across Skylake and Sandy Bridge.}
        \label{flagimpact}
\end{figure}

\begin{figure*}%
    \centering
    \subfloat{{\includegraphics[scale=0.29]{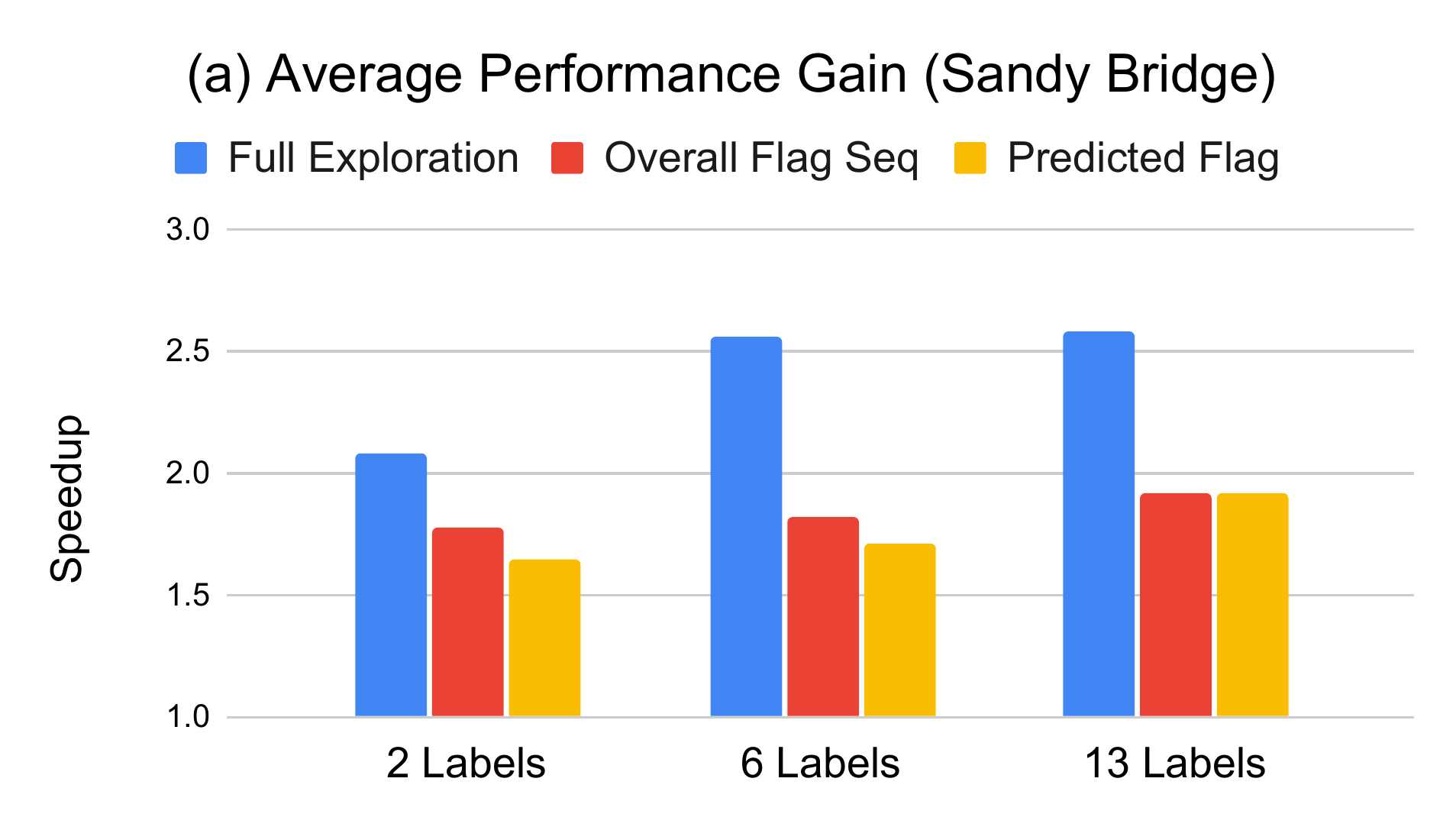} }}%
    \addtocounter{subfigure}{-1}
    \subfloat{{\includegraphics[scale=0.29]{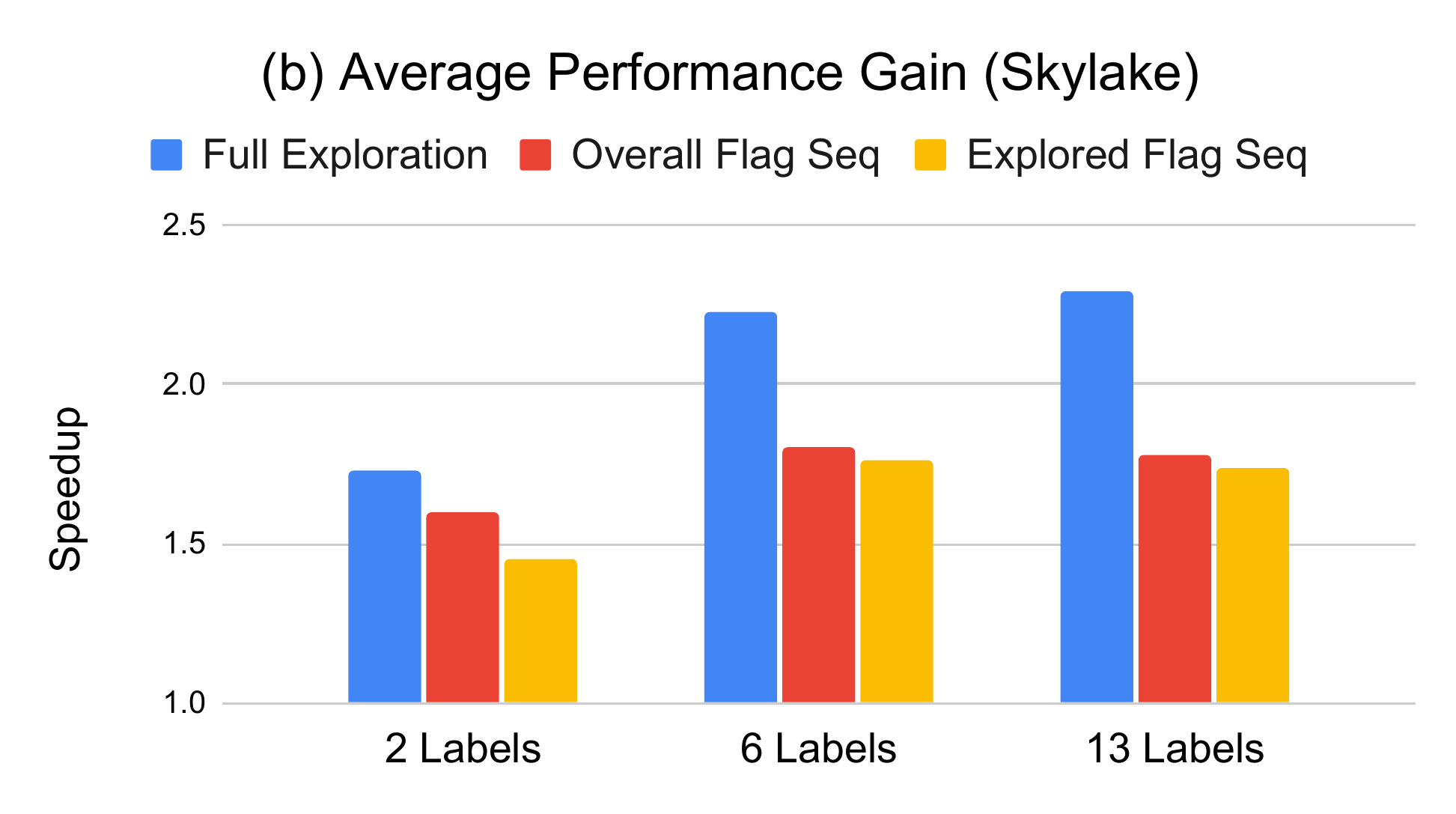} }}%
    \addtocounter{subfigure}{-1}
    \qquad
    \subfloat{{\includegraphics[scale=0.29]{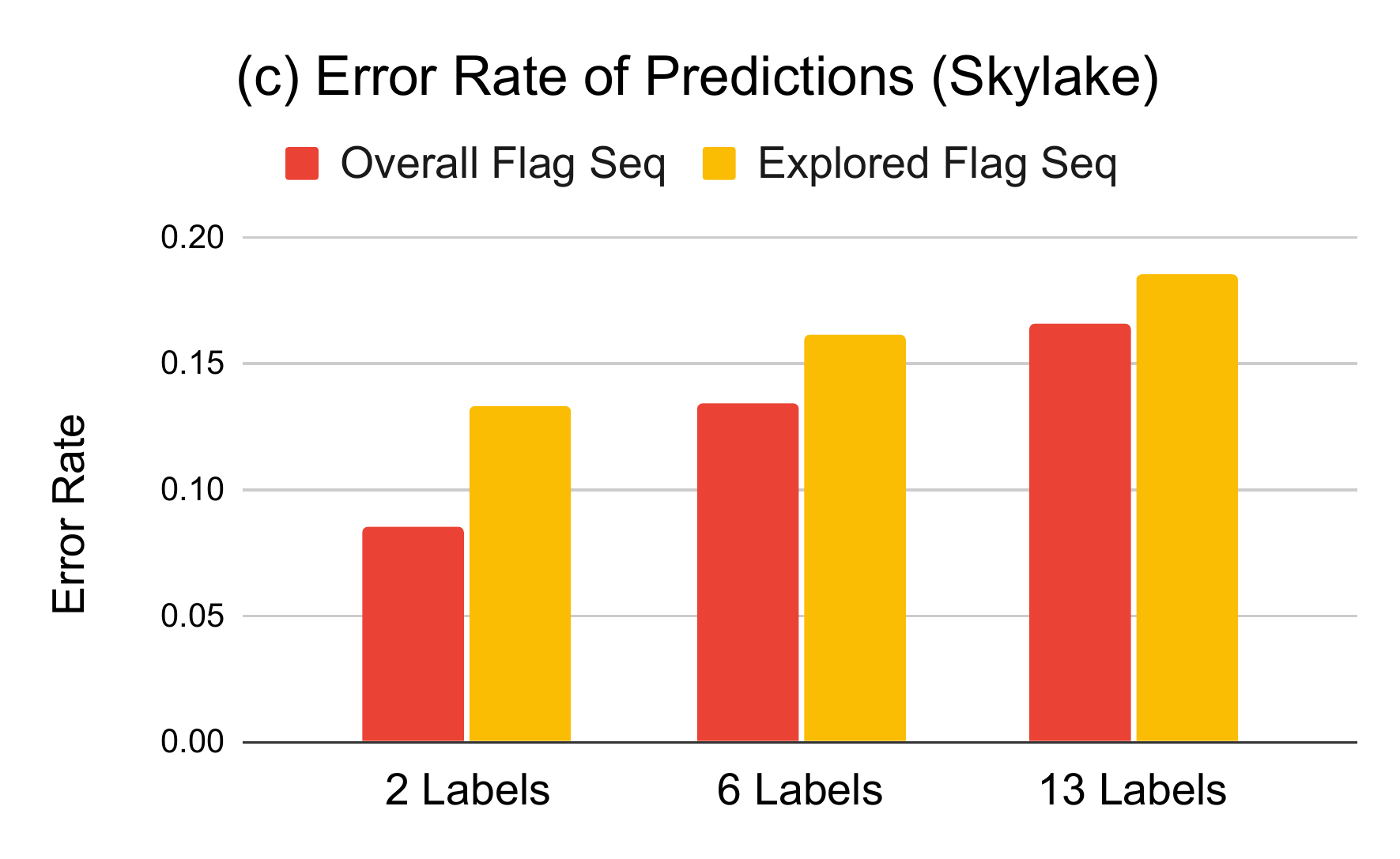} }}%
    \addtocounter{subfigure}{-1}
    \caption{[higher is better] Left and middle show performance gains of the model while varying the labels. Right presents the prediction accuracy of the models. Reducing the number of labels increases the accuracy of the model but limits its gains. }%
    \label{lesslabels}%
\end{figure*}

\subsection{Static Prediction of Configurations}
\label{sec:exp:stat}

This section evaluates the static prediction model described in Section~\ref{methodology}. We use 1000 different compiler sequences to train the static model. We consistently use the same 1000 sequences across all the training folds. However, we note that different validation folds can have different flag sequences associated with their optimization model as the training programs change across validation folds. We start by using the flag exploration method described in Section \ref{sec:stepe} and eventually evaluate the benefits of the more complex prediction flag model in Section \ref{sec:res:modelflag}. For the rest of this section, we refer to the static optimization model using the explored flag seq as \textit{static}.


Figure~\ref{errorperregion} shows for each region the prediction errors of the static and dynamic models against the full exploration (i.e., best configuration found in step C of Fig \ref{fig:training}). Errors are calculated as the relative differences (i.e., absolute difference divided by the maximum absolute value of the two numbers) between the full exploration times and the statically (blue) and dynamically (red) predicted configurations times.


Half of the regions are perfectly optimized statically.
In some cases, the static approach even predicts optimum configurations whereas the dynamic approach fails (right side of the sub-figures). However, as expected due to dynamic behavior, the static method might lack the necessary information for some regions to correctly predict the configurations (left side).

We further investigate the miss-predicted regions' distribution across the folds. If most of the miss-predictions occur within the same folds, it means that the associated training regions lack the relevant information to optimize the validation regions. However, Figure \ref{errorfold} shows that the error distribution is mostly spread across the folds evenly. 




We also explore how important flag sequences selection is for our model. Figure~\ref{flagimpact} displays the average prediction gains over both Sandy Bridge and Skylake across the 1000 flags sequences. Selecting flag sequences is important as they affect the speedup from $1.6\times$ to $1.9\times$ on Sandy Bridge. Moreover, different flag sequences are required to achieve good results depending on the micro-architecture (i.e., Sandy Bridge and Skylake need different flag sequences). While the simple exploration method finds efficient flags sequences, we observe that some flag sequences systematically outperform on Skylake. We further investigate how to optimize the flag selection process with a model in Section \ref{sec:res:modelflag}.





\subsection{Improving Predictions by Reducing the Number of Labels}
We explore the impact of varying the number of labels. We select 2 and 6 labels by minimizing their numbers while maximizing their gains as described by \cite{10.1145/3392717.3392765}.
Reducing the number of labels eases the burden on our model
but also reduces the overall potential gains.
In Figure~\ref{lesslabels}, the left and middle plots present the arithmetic average gains across different number of labels for each micro-architecture. We additionally compare the gains of the explored flag seq against
\textit{overall flag seq} (the static model predictions if it picks the most efficient flag sequence on average across all the regions independently from training and validation) and \textit{full exploration} (select the best configuration among the 13 labels for each region).

We observe that the prediction gains of the model are slightly decreased. However, our model is more accurate at predicting the best configuration, reaching less than $10 \%$ errors with the overall flag seq (see Figure \ref{lesslabels} right side). Unfortunately, the additional model prediction capabilities are compensated by the reduced potential performance gains. 

\begin{figure}
\centering
\includegraphics[width=0.5\textwidth]{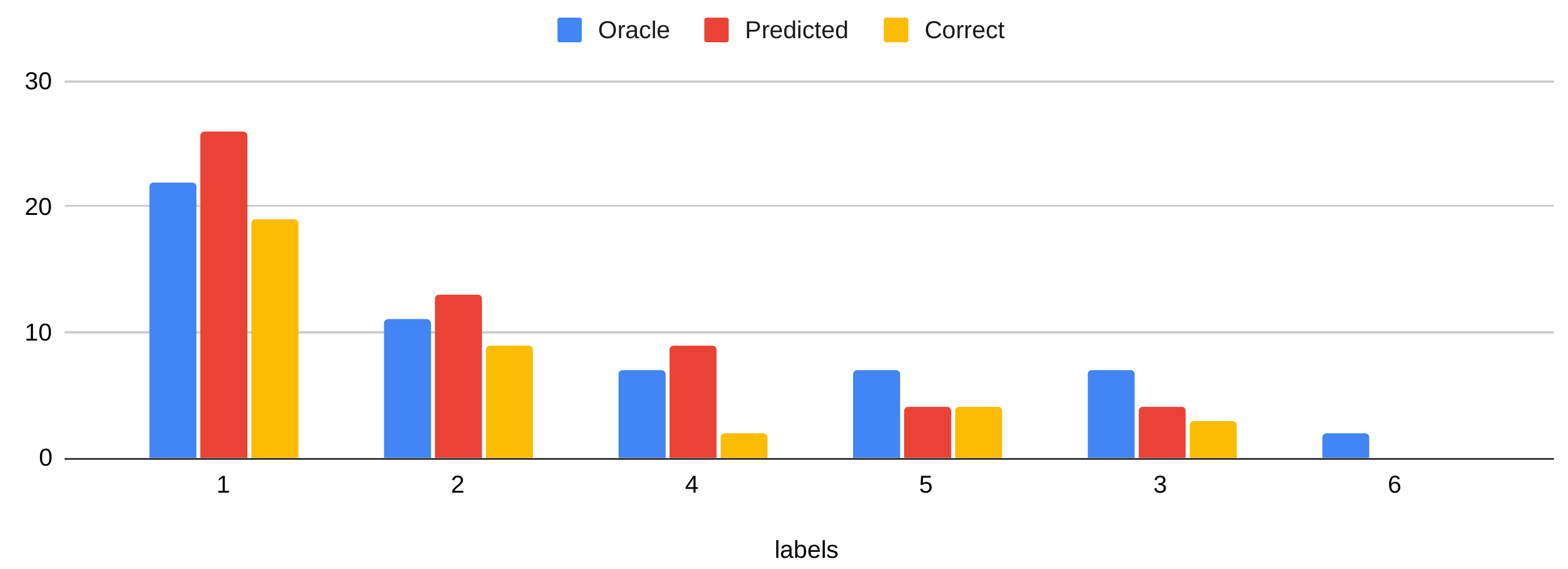}
\caption{Number of predictions per label for Skylake natively trained using 6 labels. \textit{Oracle} refers to the number of times a label is marked for a code. \textit{Predicted} presents the number of times our model predicted the label. \textit{Correct} shows how many of these predictions were correct.}
\label{label-error}
\end{figure}

To better understand these prediction errors, we present in Figure~\ref{label-error} the predictions per label for Skylake with 6 labels. Because some labels have very little instances (e.g. label 6 occurs 2 times as it represents a configuration rarely optimal), they are difficult to correctly predict. Adding more codes could alleviate this issue. Nevertheless, the miss-predictions are mostly correlated with the oracle numbers indicating that there is no systematic issue with a label.

\subsection{Cross-Architecture Prediction}
We also investigate cross micro-architecture prediction (i.e., training on Sandy Bridge and validating on Skylake and vice versa).
To do so, we trained our model for one system and then applied it to the other one by just translating the labels. Because Sandy Bridge and Skylake share the same prefetch settings, we just reapplied them as such. Similarly, we kept the thread (e.g., round robin, contiguous) and data mapping (e.g., interleave, locality) policies. However, we need to explicitly translate the number of threads and NUMA nodes to use by adjusting them to the target system. For instance, Sandy Bridge is saturated with 32 threads while Skylake is saturated with 48. To translate a thread configuration, we adjust the configuration to the available number of cores (i.e, a 48 threads configuration on Skylake is translated to a 32 threads configuration on Sandy Bridge and vice versa). We apply the same principle when the number of cores per node is different.

\begin{figure}
\centering
\includegraphics[scale=0.30]{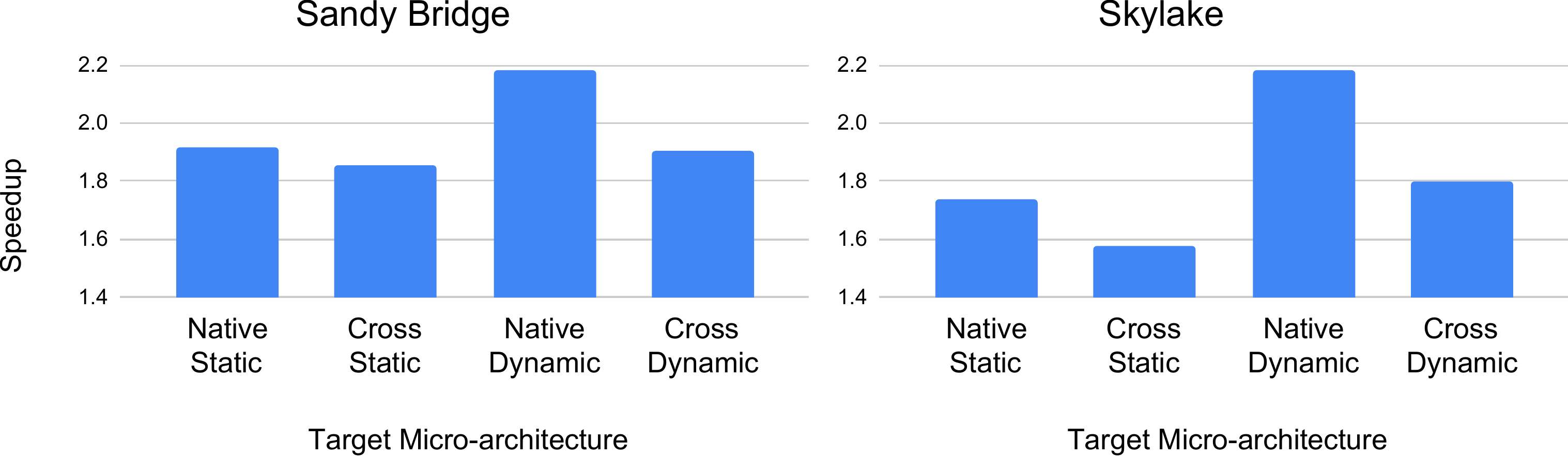} 
\caption{[higher is better] Cross architecture prediction on both Skylake and Sandy. \textit{Native} refers to training and validating on the same architecture while \textit{cross} refers to training on the other architecture and validating on the current architecture. }
\label{crosspred}
\end{figure}

Figure~\ref{crosspred} illustrates the performance gain when  training on the same micro-architecture (native) or on different micro-architecture (cross).
While cross prediction loses some of the gains due to both 1) the configuration translation and 2) the model trained on a different system, it is still able to provide an average arithmetic gains of $1.7\times$. We also cross-evaluated the dynamic model by collecting the performance counters selected for Sandy Bridge on Skylake and vice versa.  We observe that the native static model is on par with the cross dynamic one and is less impacted by cross prediction.

\begin{figure*}
\centering
\includegraphics[width=\textwidth]{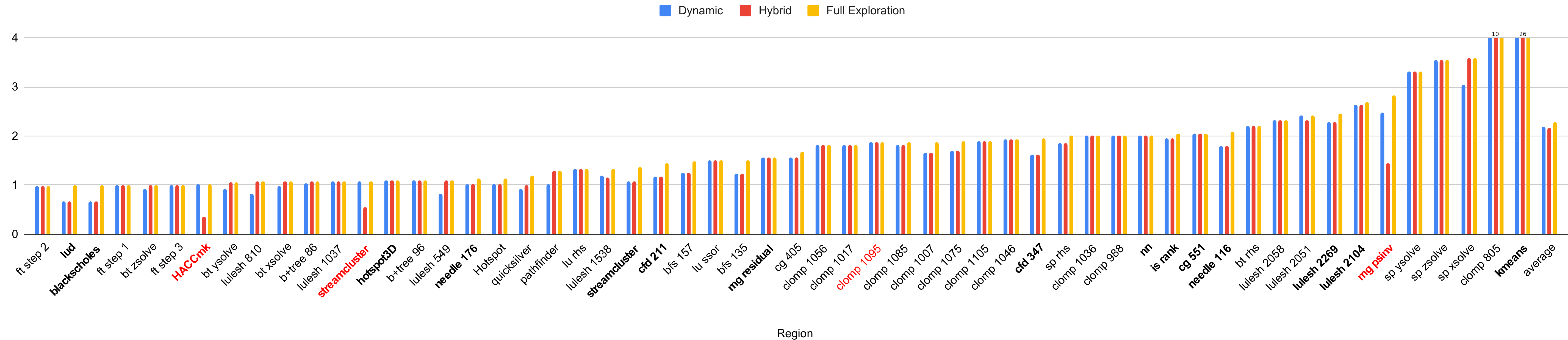}
\caption{[higher is better] Performance gains per region. Blue and red bars respectively show the performance gain achieved by the dynamic approach and the hybrid model.
Yellow is the full exploration across the entire space (each region is executed with the most efficient configuration). Regions names in bold indicate that they need to be profiled (the error of the model is higher than 20\%). Regions names in red indicate that the hybrid model failed to predict if program needs dynamic or static information. Overall, the hybrid model achieves the same gains as the dynamic one but only profiles 16 programs.}
\label{finalgains}
\end{figure*}

\subsection{Changing Input Sizes}
\begin{figure}[h]
\centering
     \begin{subfigure}{0.5\textwidth}
         \centering
         \includegraphics[width=\textwidth]{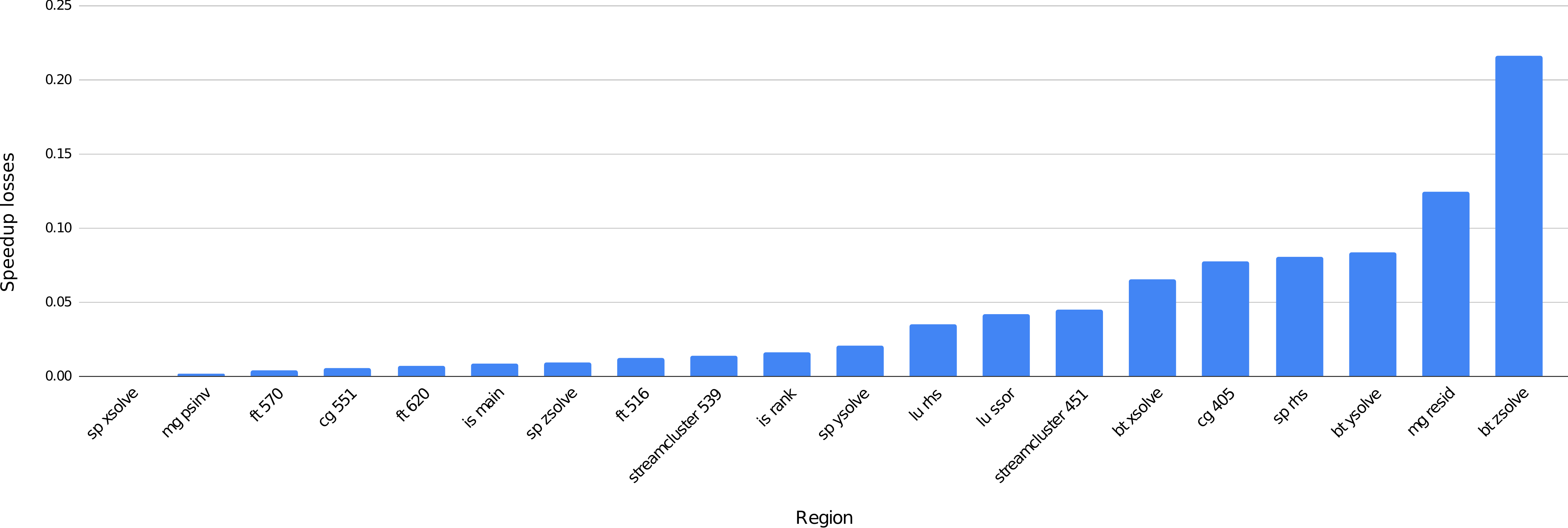}         
     \end{subfigure}
        \caption{[lower is better] Speedup losses per region with size-1 inputs. For each region, we use size-2 to optimize the regions and compare the resulting timing against a native size-1 optimization timing. We calculate the speedup losses per region as: $ L = S^{on\ size\ 1}_{with\ best\ conf\ size
1} - S^{on\ size\ 1}_{with\ best\ conf\  size2} $.}
        \label{changing_input_sizes}
\end{figure}

Because of static information, we predict a single configuration per program. However, programs behavior might change depending on their inputs, resulting in sub-optimal predictions across different inputs. To evaluate the generalization and resiliency of our method, we quantify how much changing inputs affects the best configurations and in turn reduces the performance gains of our model.

Due to system constrains, the experiments in this subsection were conducted on a Skylake (Xeon Gold 6130) from the Grid'5000 environment~\cite{grid5000} with Clang 6 and only evaluate a subset of the benchmarks. Due to the huge search space size (over 300 points), we reproduced the sampled execution by only executing $10$ calls per region\cite{10.1145/3392717.3392765}. 

We executed the regions across the entire space over 2 inputs called size-1 and size-2. Size-1 refers to CLASS A and small inputs for NAS and Rodinia respectively while size-2 refers to CLASS B and largest inputs. Natively optimizing size-1 provides an average $1.51\times$ speedup. On the other hand, optimizing each region for size-2, and then reapplying the same configurations for size-1 provides $1.46\times$ speedup, resulting in a $0.05\times$ performance gains losses due to the input changes. 
The performance losses are very region dependent suggesting opportunities to predict if a region has a consistent best configuration across different input sizes. The input impact on performance is an additional dimension to the code optimization that we further discuss in Section \ref{sec:disc}.



\subsection{Using Static or Dynamic Features}
\label{sec:exp:hyb}

As illustrated in Figure~\ref{errorperregion}, there are programs that benefit more from a static prediction than from a dynamic one. This motivates us to investigate a new hybrid solution that  enables to both reduce the cost of existing dynamic models but also to increase the gains of the static ones. 

To perform this evaluation, we implemented the hybrid model described in Section~\ref{modelhh}. In particular, we used 20\% as threshold to determine if the static model is correctly predicting the optimization for a code.

Figure \ref{finalgains} compares the hybrid approach on Skylake against the dynamic model as well as a full exhaustive exploration.
Overall, the DT model has 92\% accuracy at identifying the correct model to be used (static or dynamic).
As a result, the hybrid model outperforms the static one and is cheaper to apply compared to the dynamic one. It only profiles 30\% of the programs.
Finally, we compared the collection cost of static versus dynamic features by measuring the compilation times versus the execution times of some regions. For small programs (CG), the compilation time is similar to the execution time. However, as expected medium/large programs (SP) take order of magnitude longer to execute than to compile.
\subsection{Selecting Flag Sequences}
\label{sec:res:modelflag}

\begin{figure}
\centering
\includegraphics[scale=0.32]{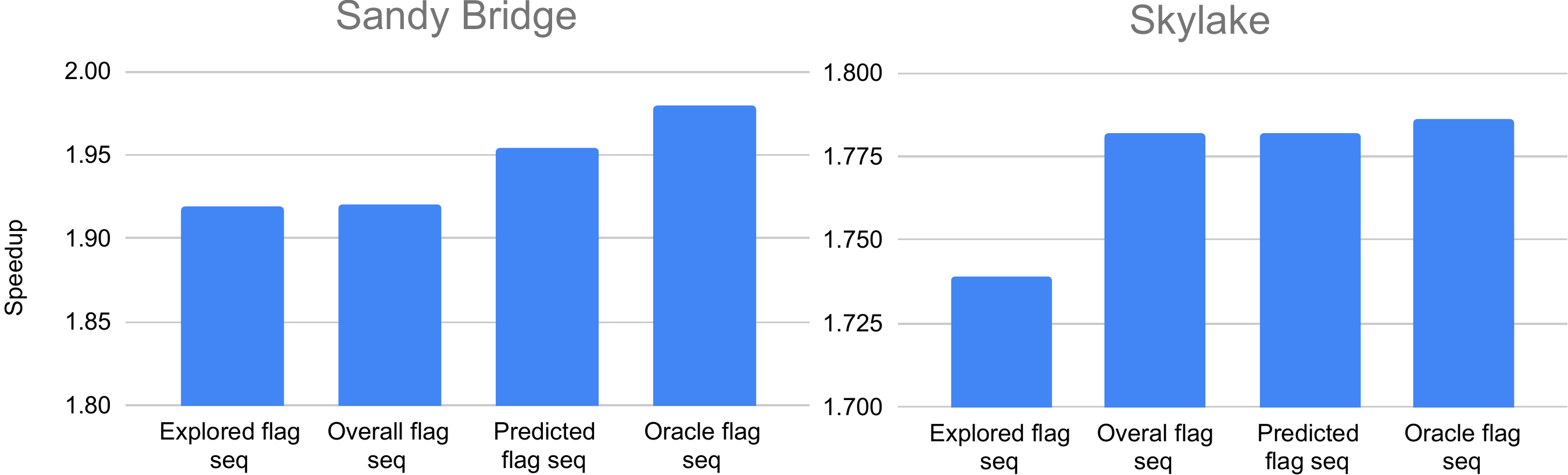}
\caption{[higher is better] Average speedup across all regions from the 10 validation folds using different flag-sequences.}
\label{flag_sequence_learner}
\end{figure}

Figure \ref{flagimpact} quantifies how important flag sequences are to achieve high performance gains. While generally efficient, we observe that the explored flag seq (single best flag sequence from training regions) method is outperformed by the overall flag seq (single best flag sequence from training and validating regions) on Skylake. Even worse, we have no guaranty that using a single flag sequence for a model is always sufficient for all the programs.

Therefore we implemented a flag prediction decision tree model described in Section \ref{sec:stepe} that adjusts the compiler sequence to the target program. The model always compiles the application using a fixed flag sequence (extracted from the 1000s explored) and use the resulting vector to predict the appropriate flag sequence for the target program (also called predicted flag seq). The model is labeled over the training regions with different flag sequences.  
To select the labels, we applied the same search method as the one selecting the 13 labels given in~\cite{10.1145/3392717.3392765}. We investigated efficient flag sequences and gathered them into lists. 2 and 4 sequences were necessary on Skylake and Sandy Bridge respectively to reach 99\% of the \textit{oracle flag seq} (the theoretical static model predictions if we pick the most efficient flag sequence per region) gains. 

Figure~\ref{flag_sequence_learner} compares the different flag selection strategies. We observe that the static performance predictions are improved by $3.4\%$ and $4.2\%$ on the two systems. Optimizing Sandy Bridge requires using different flag-sequences with the model while just picking the best flag-sequence improves performance on Skylake. 

\section{Discussion}
\label{sec:disc}

\begin{figure*}
\centering
\includegraphics[width=\textwidth]{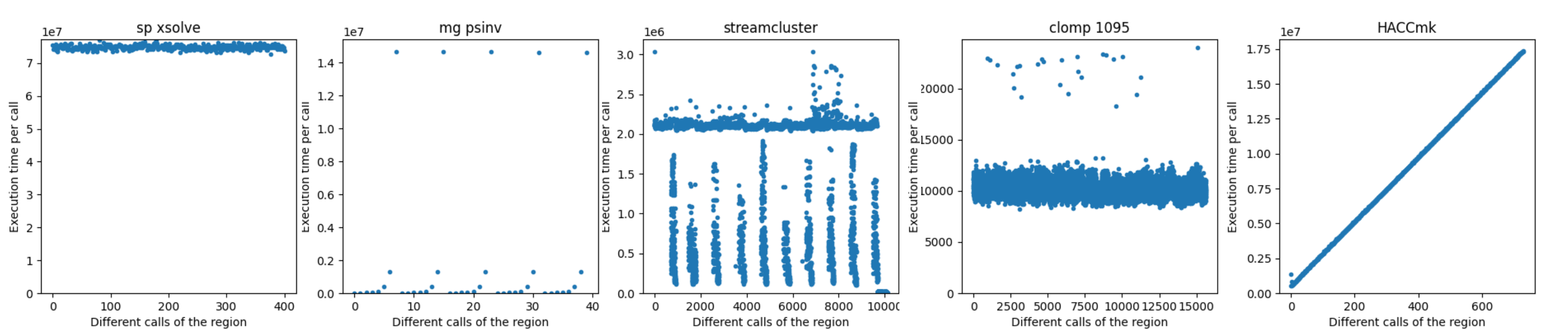}
\caption{Execution time per call (in cycles) of the 4 miss-predicted regions along SP as reference (with clang 6 Xeon Gold).}
\label{runtime_behavior}
\end{figure*}


While our experiments show that it is possible to predict the appropriate configuration for a majority of programs statically (up to 80\% of performance gains can be achieved statically), there are cases where static strategies pathologically could not predict the optimal configuration.

This is likely due to dynamic behaviors at runtime that appear based on the region dataset. We think that static information is intrinsically limited to characterize some of these behaviors. Figure \ref{runtime_behavior} presents such behavior changes in mispredicted regions plus SP (a stable region for reference).
Interestingly, while static information is sometimes insufficient to select an optimization, it can detect if additional information (dynamic) is required for optimization. We plan to explore a similar approach across inputs (predict if there is a behavior change across inputs but not actually predict the change itself) to improve the performance generalization.

Currently, we propose a two step hybrid model. We believe that by tightly coupling the static and dynamic data, we have the potential to further improve the predictions. 

We focused on Intel-based systems but our augmentation method is fundamentally independent from the system and the target optimization and it is just based on the LLVM IR. Applying our workflow to a new architecture however requires translating the NUMA/prefetching optimization space. Similarly, we can use our workflow to explore other optimizations such as compiler flags or scheduling on heterogeneous devices. 

Finally, we assume that applications are run alone on the system. Co-executing applications will change their best configurations due to contention over shared resources. We can extend our method to support such environments by exploring the labels while co-executing the applications.

%% file: 5_related_works.tex
\section{Related Works} 
\label{related}

This paper proposes a deep learning characterization method for search space optimization of NUMA and Prefetcher configurations. Broadly, our work can be categorized into 1) NUMA/Prefetcher optimization and 2) search space reduction and optimization.


While hardware prefetching aims to reduce latency, it can hurt performance by increasing cache evictions as well as interference across threads. To address these issues, adaptive prefetching schemes~\cite{wu2011characterization,khan2015arep,jimenez2012making} directly execute different configurations to identify the most efficient ones. Similarly, with the increased adoption of NUMA architectures, placement of threads and data for parallel applications is of utmost importance\cite{popov2019efficient} due to the resulting remote accesses and congestion on memory controllers. Radojkovic et al.\cite{radojkovic2016thread} execute different combinations of thread mappings to improve performance. Popov et al.~\cite{popov2019efficient} increased the performance gains by further expanding the search with data mappings.




Exploration driven strategies successfully optimize NUMA and prefetching but require executing hundreds of configurations. To avoid such overhead, machine learning strategies build models to predict efficient configurations based on features that can be collected without doing the full search. Liao et al.~\cite{liao2009machine} collect cache (i.e., L1, L2, and DTLB) and bus related performance counters to optimize the prefetch configurations with decision trees. Hiebel et al.\cite{Hiebel:2019} further monitors off cores requests and branch miss predictions to adjust the prefetching configuration at runtime according to the dynamic workload performance behavior.  Wang et al.~\cite{wang2009mapping} predict the number of threads along with the scheduling policies to use. Castro et al.~\cite{castro2011machine} model the thread scheduling strategy (e.g., scatter, round robin). Wang et al.~\cite{wang2016predicting} further select the right number of NUMA nodes to use for thread mapping. Denoyelle et al.~\cite{Denoyelle:2019:DTP:3337821.3337893} complement this exploration with data placement. 
Entezari et al.~\cite{entezari2020evaluation} thoroughly characterize NUMA memory behaviors with Stochastic Reward Nets while Sánchez Barrera et al.~\cite{10.1145/3392717.3392765} extend these explorations by coupling both NUMA thread and data configurations with prefetching. Diener et al.~\cite{diener2017affinity} and Mittal et al.~\cite{mittal2016survey} provide more context and details about these broad areas in 2017 and 2016 surveys on NUMA and prefetching respectively. 




More generally, classical approaches to compiler and search space optimizations rely on heuristic cost functions to estimate their quality \cite{wang2018machine}.  
Qilin \cite{luk2009qilin} uses adaptive mapping to offload computations to the CPU/GPU. 
In \cite{stephenson2003meta}, Stephenson et al. introduce Meta Optimization, a machine learning based approach for tuning heuristics to increase speedups. Hoste et al.\cite{hoste2008cole} propose the COLE framework to find efficient compiler optimization levels using multi-objective evolutionary search. 
Cavazos et al.\cite{cavazos2006method} use logistic regression for determining the optimization level of Jike RVM. 

To the best of our knowledge, our strategy is the first to optimize the complex space of NUMA and prefetching with only static information. It is inherently limited by the absence of some dynamic behaviors but can be cheaply applied without any performance profiling. Furthermore, we reduce the cost of dynamic optimization strategies by only identifying cases where the static information was insufficient.


%% file: 6_conclusion.tex
\section{Conclusion}
\label{conclusion}
An appropriate configuration of NUMA and hardware prefetchers has a huge impact on the performance gain of programs.
Given the huge tuning parameter search space, selecting the right configuration is a non-trivial task.
Numerous efforts have been proposed to automate the process of identifying the correct configurations, however the majority of these approaches rely on dynamic properties of applications which are very costly to acquire.
In this paper, we propose a novel approach based on graph neural networks that predicts the optimal configuration by relying on the static features of applications. While we could achieve 80\% of performance gain of dynamic approaches, the static approach could not predict the optimal configurations for a few cases. 
To alleviate those cases, we further developed a hybrid approach when dynamic features are required.
Our results show that the hybrid approach is competitive with state-of-the-art dynamic approaches at lower profiling cost and resource consumption.

\section*{Acknowledgments}

We would like to thank the Research IT team\footnote{https://researchit.las.iastate.edu} of Iowa State University for their continuous support in providing access to HPC clusters for conducting the experiments of this research project. Experiments presented in this paper were carried out using the experimental testbeds PlaFRIM, supported by Inria, CNRS (LABRI and IMB), Université de Bordeaux, Bordeaux INP, Conseil Régional d’Aquitaine\footnote{https://www.plafrim.fr}, and Grid'5000, supported by a scientific interest group hosted by Inria, 
CNRS, RENATER and several Universities as well as other organizations\footnote{https://www.grid5000.fr}.

